\documentstyle[twoside,fleqn,espcrc2,epsfig,physics]{article}

\newcommand{\beq}{\begin{equation}}
\newcommand{\eeq}{\end{equation}  }
\newcommand{\bec}{\begin{center}}
\newcommand{\eec}{\end{center}}
\def\vs{\vskip}

\def\e{\epsilon}
\def\as{\alpha_s}

\def\Z{\mbox{Z}}

\def\Th{\Theta}
\def\Th0{\Theta_0}
\title{Multiplicities, fluctuations and QCD: Interplay between soft and hard
physics?
\thanks{Presented at the XXVII Symposium on Multiparticle Dynamics,
September 8-12, 1997 Frascati-Rome, Italy.}}

\author{W. Kittel with S.V. Chekanov
\thanks{On leave from
Institute of Physics,  AS of Belarus,
Skaryna av.70, Minsk 220072, Belarus.},
D.J. Mangeol and W.J. Metzger
\vs 2mm
        High Energy Physics Institute Nijmegen (HEFIN), 
        University of Nijmegen/NIKHEF,
        P.O. Box 9010, 6500 GL Nijmegen, The Netherlands
\vs 2mm
for the L3 Collaboration
}
\begin{document}

\begin{abstract}
Multiplicity fluctuations are studied both globaly (in terms of high-order
moments) and locally (in terms of small phase-space intervals). The ratio of 
cumulant factorial to factorial moments of the charged-particle multiplicity 
distribution shows a quasi-oscillatory behaviour similar to that predicted by 
the NNLLA of perturbative QCD. However, an analysis of the sub-jet 
multiplicity distribution at perturbative scales shows that these 
oscillations cannot be related to the NNLLA prediction. We investigate how 
it is possible to reproduce the oscillations within the framework of 
Monte-Carlo models. Furthermore, local multiplicity fluctuations in angular 
phase-space intervals are compared with Monte-Carlo models and with 
first-order QCD predictions.  While JETSET reproduces the experimental data 
very well, the predictions of the Double Leading Log Approximations and 
estimates obtained in  Modified Leading Log Approximations deviate 
significantly from the data.
\end{abstract}

\maketitle

\section{INTRODUCTION}

The shape of a distribution can be analyzed in terms of its moments 
\cite{kittel}. In the case of the multiplicity of particles produced 
in particle collisions at high energies, such an approach is prefered 
both theoretically (generating function) and experimentally (stability). 

The normalised factorial moment of rank $q$ 
of the multiplicity distribution $P_n$ is defined by:

\beq
F_q=\frac{\sum^\infty_{n=q}n(n-1)....(n-q+1)P_n}
 {\left (\sum^\infty_{n=1}nP_n\right )^{q}}\ .
\eeq
It expresses the normalized phase-space integral over the $q$-particle
density function. If particles are produced independently, the multiplicity
distribution is a Poissonian (for which the factorial moments of all ranks
are equal to one). If the particles are positively
correlated, the distribution is 
broader than Poisson and the factorial moments are greater than one.

The cumulant factorial moments are obtained from the factorial moments
by:
\beq
K_q=F_q-\sum^{q-1}_{m=0}\frac{(q-1)!}{m!(q-m-1)!} K_{q-m}F_m \ .
\eeq
$K_q$ expresses the normalized phase-space integral over the $q$-particle 
correlation function, i.e., that function which describes the genuine 
correlations between $q$ particles. 

Here, we shall use these moments, for full phase space, as well as in
ever smaller intervals of it, to study the interplay between hard and
soft physics in particle production. In doing so, we use perturbative
QCD (and its anomalous dimension) not as a model to be tested, but as
a description of the perturbative background at the onset of unknown
non-perturbative hadronization phenomena.

\section{THE GLOBAL MULTIPLICITY DISTRIBUTION}

\subsection{High-order correlations}

Since $|K_q|$ and $F_q$ increase rapidly with increasing $q$, it
is useful to define \cite{dremin2} the ratio $H_q=K_q/F_q$, 
which has the same order of magnitude over a large range of $q$. One 
can view this ratio as a cumulant factorial moment normalised to the 
factorial moment. It reflects the genuine $q$-particle correlation 
relative to the $q$-particle density.

\begin{figure}[tp]
\epsfig{file=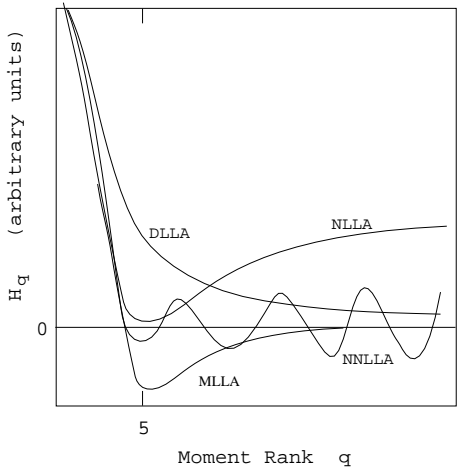,width=6cm}
\vs-7mm
  \caption{Predictions at different order of perturbative QCD for the
  $H_q$ as a function of the rank $q$ [2].}  \label{fig:pred}
\vs-6mm
\end{figure}

The $H_q$ have been calculated for the gluon multiplicity distribution 
at different orders of perturbative QCD \cite{dremin2} (see Fig.~1):

{\leftskip=5mm
\hskip-7mm
$\bullet$ For the Double Leading Logarithm Approximation (DLLA), $H_q$
decreases to 0 as $q^{-2}$.

\hskip-7mm 
$\bullet$ For the Modified Leading Logarithm Approximation (MLLA), $H_q$
decreases to a negative minimum at $q=5$, and then rises to approach 0
asymptotically.

\hskip-7mm 
$\bullet$ For the Next to Leading Logarithm Approximation (NLLA), $H_q$
decreases to a positive minimum at $q=5$ and goes to a positive
constant value for large $q$.

\hskip-7mm 
$\bullet$ For the Next to Next to Leading Logarithm Approximation (NNLLA),
$H_q$ decreases to a negative first minimum for $q=5$ and shows 
quasi-oscillation around 0 for $q>5$.
\par}

Assuming the validity of the Local Parton Hadron Duality hypothesis
(LPHD), such behaviour is also expected for the charged-particle
multiplicity distribution.

\begin{figure}[tp]
\vs1mm
\epsfig{file=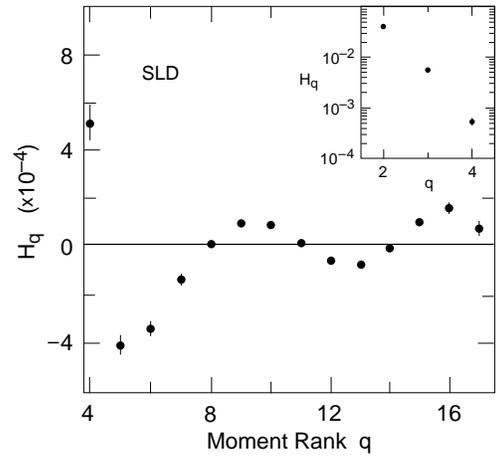,width=6.5cm}
\vs-7mm
  \caption{$H_q$ versus $q$ measured by the SLD collaboration [3].}
  \label{fig:SLD} 
\vs-6mm
\end{figure}

The $H_q$ of the charged-particle multiplicity distribution have been
measured by the SLD collaboration~\cite{SLD} for e$^+$e$^-\longrightarrow 
\mathrm{hadrons}$ at $\sqrt{s}=M_{\mathrm{Z}}$. They
indeed observe (Fig.~\ref{fig:SLD}) a negative minimum for $q=5$ followed
by quasi-oscillations about 0 for higher $q$.  This result has been
interpreted as confirmation of the NNLLA prediction.

Earlier, the $H_q$ moments were calculated \cite{drem94,shoulder} from charged
particle multiplicity distributions of e$^+$e$^-$ experiments between 22
and 91 GeV. Similar behaviour with a
negative first minimum followed by quasi-oscillation about 0 is obtained.
Furthermore, the same behaviour has been observed 
in hadron collisions between 20 and $900\GeV$ \cite{drem94,hqhadron} and
in hA an AA collisions \cite{capel97} .

While it is tempting to conclude that the observed behaviour is a
confirmation of the NNLLA prediction, it must be realized that such a
conclusion rests strongly on the validity of LPHD.  Although the LPHD
hypothesis is accepted for single-particle distributions and mean
multiplicity, difficulties have been detected when resonances are
involved~\cite{delphi_lphd,opal_lphd}.  Further, heavy quark decays
will cause a modification of the multiplicity distribution which is
unlikely to be accounted by LPHD~\cite{shoulder}. For these reasons, we
consider the assumption of LPHD as highly questionable.

Therefore, we present measurements of $H_q$ not only of the charged-particle
multiplicity distribution but also of the sub-jet multiplicity
distributions.  By considering sub-jets, i.e., jets obtained with
small value of jet resolution parameter, we can, in some sense, follow
the development of parton to hadron by varying the jet resolution
parameter (and hence the energy scale).  The variation of the mean
sub-jet multiplicity has been found~\cite{subj_opal} to agree well with
the analytic perturbative QCD calculation in the perturbative region
(energy scale greater than about $1-2\GeV$) and with Monte Carlo in
both the perturbative and non-perturbative region.  These sub-jet
multiplicity distributions should therefore enable us to extend this
analysis to the energy scales of partons and to that of perturbative
QCD, avoiding (or at least lessening) dependence on the assumption of
LPHD.

\subsection{The data}

The analysis is based on data collected by the L3 detector~\cite{subl3} 
at the energy of $\mathrm{Z}$, corresponding to approximately 
1.5M hadronic $\mathrm{Z}$ decays. Events are selected using 
a procedure based on the track momenta measured in the Central Tracking 
Detector including the Silicon Micro-vertex Detector.
About 1M  events survive this selection.

The charged-particle multiplicity distribution was
corrected for the detector acceptance and efficiency using a
 matrix~\cite{opal_44} determined from Monte-Carlo events generated by
 JETSET~\cite{sjo} and passed through detector simulation.  
Effects from selection procedure and initial state radiation
are corrected from detector simulation. Charged particles originating from
$\mathrm{K^{0}_{S}}$ and $\Lambda$ decay are left in the sample.

The same procedure was applied to correct the sub-jet multiplicity 
distributions.

To evaluate the errors of the $H_q$, we generated 10000 multiplicity 
distributions from the experimental one, by allowing a Gaussian fluctuation 
of each $P(n)$. The widths of the Gaussians were given by the statistical 
errors of the $P(n)$. The statistical error on the $H_q$ is taken as the 
standard deviation of the 10000 values of $H_q$. 
    
\subsection{Charged-particle multiplicity analysis}
\begin{figure}[tb] 
\vs-1.3truecm
\epsfig{file=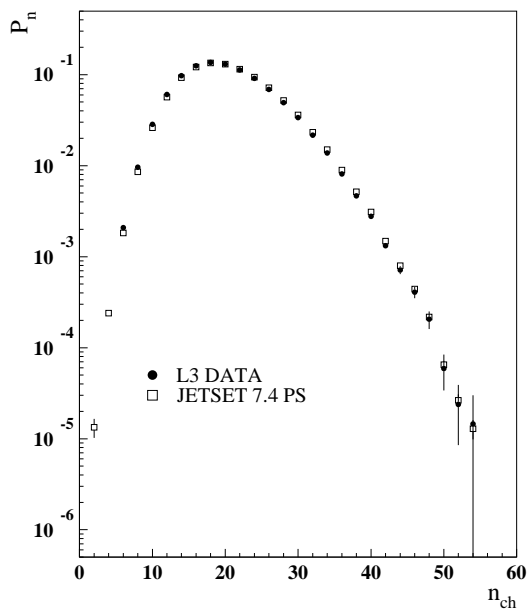,width=8cm}
\vs-8mm
\caption{The fully corrected multiplicity
    distribution compared to JETSET 7.4 PS.}
    \label{fig:cor}
\vs-5mm
\end{figure}

The detector-corrected charged-particle multiplicity distribution is 
shown in Fig.~\ref{fig:cor} and compared 
to that of JETSET. Reasonably good agreement is found.

\begin{figure}[tb]
\vs-0.5truecm
\epsfig{file=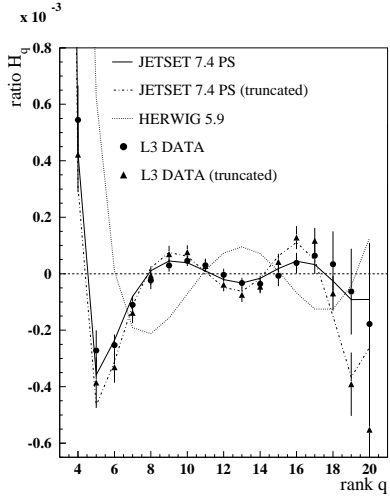,width=6cm}
\vs-10mm
  \caption{$H_q$ obtained from the charged-particle multiplicity distribution.}
  \label{fig:hq}
\vs-5mm
\end{figure}

To be less sensitive to the large statistical fluctuations in the 
high-multiplicity tail due to a finite data sample, the largest 
multiplicities ($N_{\mathrm{ch}}>50$) are removed. The ratios $H_q$ 
are shown in Fig.~\ref{fig:hq} for both the initial
charged-particle multiplicity distribution and the truncated one.
These results are in qualitatively good agreement with those of the
SLD collaboration.

In both the truncated and the non-truncated cases, the $H_q$ have a
negative first minimum at $q=5$ and quasi-oscillations around 0 for
higher $q$,  but the amplitudes are slightly larger in the truncated case
and are found to increase for further truncation (not shown).

The $H_q$ calculated from JETSET 7.4 PS agree well with the data both,
for the entire and the truncated distributions, when the generated
multiplicities are truncated in the same way as the data. The
$H_q$ calculated from HERWIG 5.9~\cite{marc} do not agree with the
data. They also show an oscillatory behaviour, but with a negative first
minimum at $q=7$ and all minima and maxima shifted by 2 units of $q$.

\subsection{Sub-jet multiplicity analysis}

Since the behaviour of the $H_q$ agrees qualitatively with that
predicted in NNLLA, it is natural to interpret this agreement as a
confirmation of the NNLLA prediction, and SLD \cite{SLD} indeed gave 
this interpretation. 
However, JETSET agrees with the data and HERWIG shows qualitatively 
similar behaviour, in spite of parton-shower modellings not being at the 
NNLLA order of pQCD and, therefore, not expected to oscillate. That they
do, together with doubt on the LPHD assumption, suggests that we
view the NNLLA interpretation with scepticism.

The sub-jet analysis assumes that the sub-jet multiplicity
distribution is related to the parton multiplicity distribution at the
energy scale corresponding to the jet resolution.  By choosing a scale
where perturbative QCD should be applicable ($\geq 1\GeV$), we
should be able to test the pQCD predictions for the behaviour of the
$H_q$ without the LPHD assumption. We define sub-jets using the Durham 
algorithm~\cite{dur}, minimizing the hadronization effects by associating
particles with large angles and small transverse energies.

It has been shown~\cite{subj_opal,subl3,alex} in previous papers (see
for review~\cite{khoze}) that sub-jet mean multiplicities agree well
with the NLLA prediction in the perturbative region, i.e., for values
of $y_{\mathrm{cut}}$ greater than $10^{-3}$ which corresponds to
transverse energies greater than $\sim 1\GeV$.

Since this study starts with the charged-particle multiplicity
distribution, we have also constructed sub-jets using only charged
tracks. The charged-particle and sub-jets multiplicities are, of
course, identical for sufficiently small values of $y_{\mathrm{cut}}$
($y_{\mathrm{cut}}< 10^{-8}$). We have verified that using 
both charged and neutral particles gives similar results.

The average sub-jet multiplicities
in 2-jet and 3-jet events ($M_2$ and $M_3$) are obtained for a set of
values of cut-off parameter $y_0$: $10^{-7} \leq  y_0 <y_1$, where
$y_1=0.01$ is the cut-off parameter which resolves 2-jet and 3-jet
events.  $M_2(y_0)$  and $M_3(y_0)$ correspond to the number of
gluons radiated at different energy scales (at different places in the
parton shower evolution) for 2-jet and 3-jet events, respectively.

\begin{figure}[tb]
\vs-0.05cm
\epsfig{file=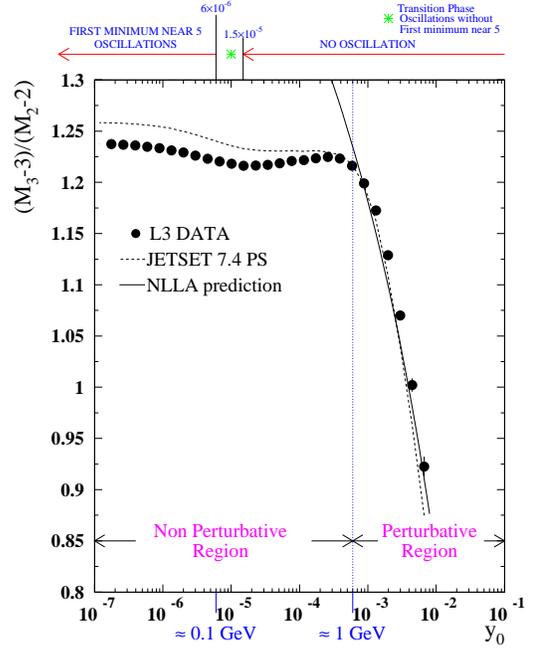,width=7.5cm}
\vs -1cm
  \caption{The soft gluon production rate as a function of $y_0$ is compared 
to the NLLA prediction. 
  Furthermore at the top of the figure, is indicated the evolution of the 
behaviour of $H_q$ with 
  the energy scales.}
  \label{fig:rat}
\vs-7mm
\end{figure}

 We have calculated the ratio $\frac{M_3-3}{M_2-2}(y_0)$, which
 corresponds to the ratio of the amount of soft gluon production, when
 the primary quarks and the hard gluon have been taken out. This
 ratio, which has the advantage of being independent of
 $\alpha_{\mathrm{s}}$, depends only on the resolution parameters
 $y_0$ and $y_1$.  The NLLA prediction \cite{webber} with $N_f=3$ \cite{nf3}
is seen (Fig.~\ref{fig:rat})
 to be in good agreement with the data in the perturbative region.
 This result supports the assumption that we can perform the sub-jet
 analysis with sub-jets obtained from charged particles.

\begin{figure}[tb]
\epsfig{file=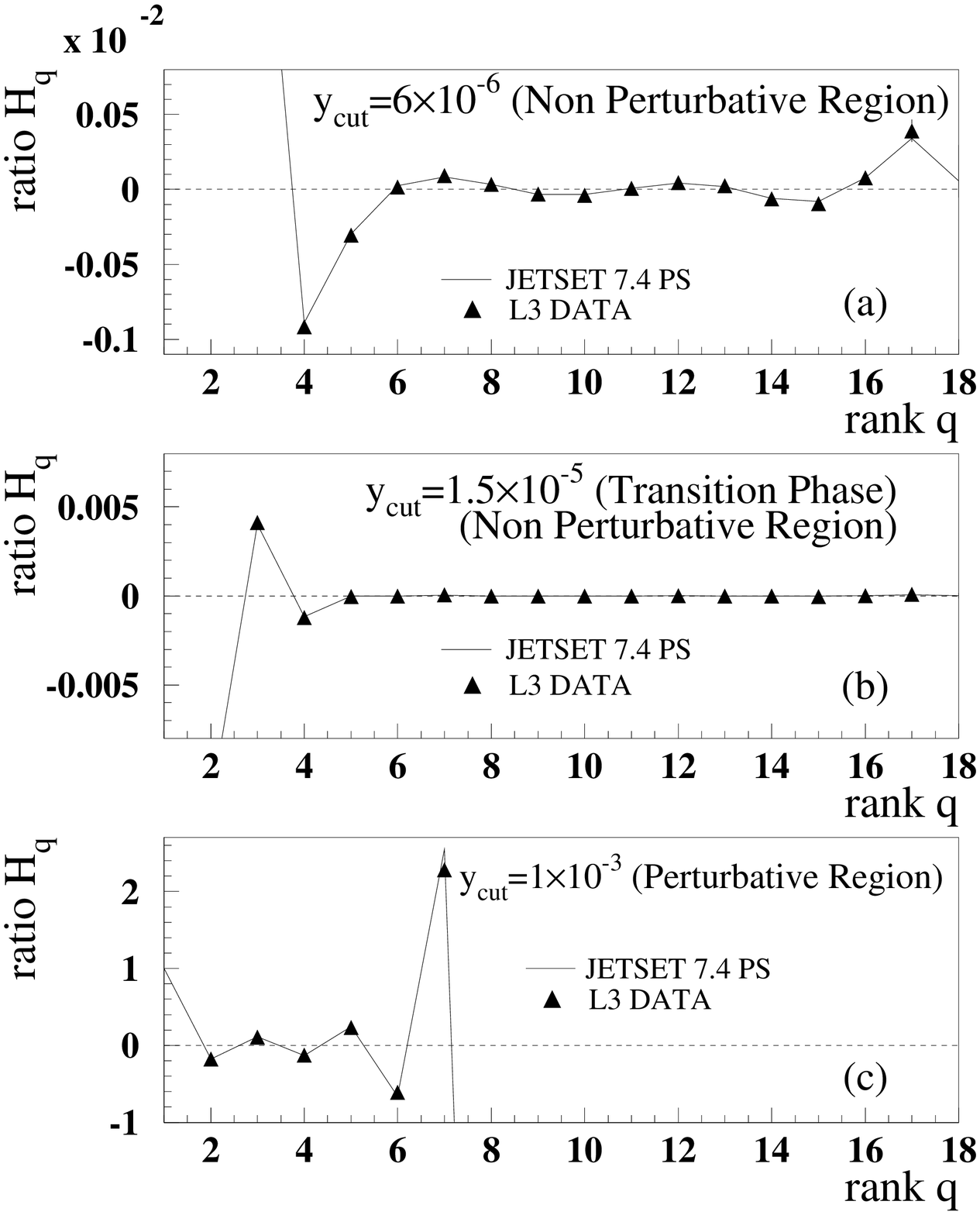,width=6cm}
\vs -1cm
  \caption{
$H_q$ for sub-jet multiplicities, (a) for $y_{\mathrm{cut}}=6\times10^{-6}$, 
(b) for $y_{\mathrm{cut}}=1.5\times10^{-5}$, (c) for $y_{\mathrm{cut}}=
1\times10^{-3}$.}
  \label{fig:sub}
\vs-8mm
\end{figure}

The sub-jet multiplicity distributions used for the calculation of the
$H_q$ have been corrected using the same procedure as used for the 
charged-particle multiplicity distribution (i.e., correction matrix and
correction for ISR, $\mathrm{K^0_S}$ and $\Lambda$ decays).  The
resulting $H_q$ are shown in Fig.~\ref{fig:sub} for three different
values of $y_{\mathrm{cut}}$.

A first minimum at $q=4$ followed by quasi-oscillation about 0, is
seen for $y_{\mathrm{cut}}= 6\times 10^{-6}$ (energy scale
$\sim  100\MeV$) (Fig.~\ref{fig:sub}a).  This behaviour is
qualitatively similar to that for the charged-particle multiplicity
distribution, as well as for sub-jet multiplicity distributions with
smaller $y_{\mathrm{cut}}$, but both the position of the first minimum
and of the subsequent maxima and minima are shifted to lower values of
$q$.

As $y_{\mathrm{cut}}$ is increased, the negative first minimum near 5
disappears. For $q\leq 5$ each $H_q$ alternates between positive
and negative values. For higher $q$, the oscillations begin to
disappear. For $y_{\mathrm{cut}}=1.5\times 10^{-5}$ (energy scale
$\sim  160\MeV$) these oscillations have disappeared
(Fig.~\ref{fig:sub}b).

In the perturbative region (energy scale $\geq  1\GeV$), the
oscillations and the negative first minimum near 5 have disappeared
(Fig.~\ref{fig:sub}c).  Instead, $H_q$ alternates between positive and
negative values for each $q$ (mind the scale!).

Although good agreement with NLLA is found for the production rates of
soft gluons seen in the perturbative region, none of the predictions
made at different orders of pQCD for the $H_q$ is observed at LEP I
energy.  Even though pQCD describes collective processes (like
the average sub-jet multiplicity) it seems to be unable to describe
the shape of these distributions.  Therefore, the minimum and
oscillatory behaviour of the $H_q$ seen for the charged-particle
multiplicity distribution and for very low values of
$y_{\mathrm{cut}}$ appears unrelated to the behaviour of the $H_q$
calculated in NNLLA.

However JETSET PS reproduces the observed behaviour at all values of
$y_{\mathrm{cut}}$, and HERWIG shows a qualitatively similar
behaviour.

\subsection{Monte Carlo study of the oscillatory behaviour of the $H_q$}

\begin{figure}[t]
\epsfig{file=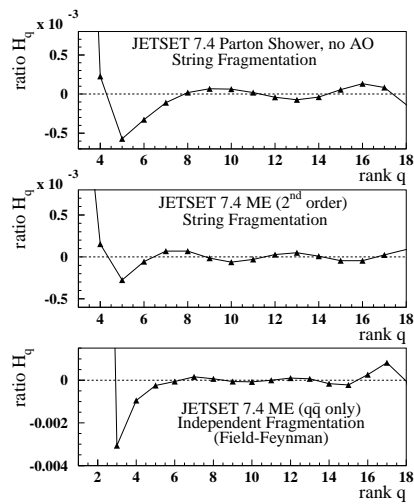,width=6cm}
\vs-1.cm
\caption{$H_q$ for the charged particle multiplicity
    distribution obtained from Monte Carlo generators using different
    methods of parton generation and different models of
    fragmentation.}  \label{fig:mc}
\vs-4mm
\end{figure}

Given the success of JETSET and HERWIG and the failure of the NNLLA
prediction seen in the previous sub-section, we attempted to find that
aspect of the Monte Carlo generators which is responsible for the
agreement.

We varied several options in JETSET and studied their influence
on the behaviour of the $H_q$.  First we used different models of
parton generation using in all cases the Lund string fragmentation:

{\leftskip=5mm
\hskip-7mm
$\bullet$ no angular ordering in the parton shower;

\hskip-7mm
$\bullet$ ${\cal{O}}(\alpha_{\mathrm{s}})$ and
${\cal{O}}(\alpha^{2}_{\mathrm{s}})$ matrix elements;

\hskip-7mm
$\bullet$ matrix element production of only ${\mathrm{q}}\bar{\mathrm{q}}$.
\par}

Next we considered the possibility that the observed behaviour could come 
from the fragmentation model. We used JETSET 7.4 with the options of
parton generation given above but with independent instead of string
fragmentation. We also used HERWIG 5.9 which uses cluster fragmentation.

In all cases the $H_q$ have a negative first minimum near $q=5$ and
quasi-oscillations for higher $q$. Some very different examples are shown in 
Fig.~\ref{fig:mc}. This Monte-Carlo study shows that one can reproduce the 
oscillatory behaviour of $H_q$ without the need for NNLLA of pQCD.

We further found that the oscillations do not exist on the parton level of
JETSET at 91 GeV, even when reducing the cut-off $Q_0$ to 0.7 GeV. They do,
however, set in at $\sqrt s=190$ GeV.

\subsection{Conclusions from the global distribution}

The ratio $H_q$ of cumulant factorial to factorial moments has been measured 
for both the charged-particle and sub-jet multiplicity distributions. A 
perturbative QCD calculation at the next-to-next-to-leading-logarithm order 
predicts $H_q$ to have a negative first minimum at $q\sim  5$ and 
quasi-oscillation around 0 for higher $q$, thus 
confirming previous results. This behaviour is indeed observed for the 
charged-particle multiplicity distribution. However, our analysis 
reveals that this behaviour appears only for very small 
$y_{\mathrm{cut}}$, corresponding to energy scales $< 100\MeV$, 
far away from the perturbative region. Furthermore, an investigation 
performed on very different models of parton generation and 
fragmentation shows similar oscillatory behaviour for all cases.

Contrary to \cite{SLD}, we conclude that the oscillations of $H_q$ 
in the charged-particle multiplicity distribution are unrelated to the 
behaviour predicted by the NNLLA perturbative QCD calculations
at the given energy.

\section{LOCAL FLUCTUATIONS}

\subsection{Introduction}

Local multiplicity 
fluctuations have been studied for many years in terms of
a variety of phase-space variables \cite{kittel},  
but  only recently has  substantial 
progress  been made to derive  analytical QCD predictions
for these observables \cite{ochs,drem,brax}. 
Assuming LPHD once again, 
these  predictions for the parton level
can be compared to experimental data.
DELPHI \cite{del} has shown that the analytical predictions tend to 
underestimate
the correlations between particles in small angular windows 
for a QCD dimensional scale of $\Lambda\sim 0.1 -0.2\GeV$. 
However, a reasonable agreement is achieved  for  an effective
$\Lambda\sim 0.04\GeV$,  surprisingly smaller than theoretical
QCD estimates \cite{pdg}. 

In an investigation \cite{l3p} of 
local multiplicity fluctuations in terms of bunching parameters \cite{bp}, 
L3 concluded that the local fluctuations inside jets are of multifractal type.
This is qualitatively consistent with  the QCD predictions. Here, we extend 
this study and present a quantitative comparison of the analytical first-order
QCD predictions \cite{ochs,drem,brax}  with the L3 data using normalized 
factorial moments of ranks $q=2$ to 5  in angular phase-space intervals. 

\subsection{Analytical  predictions}

QCD is inherently intermittent and QCD predictions \cite{ochs,drem,brax} for 
normalized factorial moments (NFMs) have the following scaling behavior 
\beq
F_q(\Theta) \propto
\left(\frac{\Theta_0}{\Theta}\right)^{(D-D_q)(q-1)}, 
\label{an1}
\eeq
where
$\Theta_0$ is the half opening angle of a cone  around the jet-axis,
$\Theta$ is the angular half-width of 
a ring around the jet-axis centered at $\Theta_0$
(see Fig.~\ref{pic}),  $D$ is the underlying topological dimension
($D=1$ for single angle $\Theta$), and $D_q$ are the
so-called R\'{e}nyi dimensions.    
The analytical QCD expectations  for  $D_q$
are as follows \cite{ochs,drem,brax}:

\begin{figure}[tb]
\vs-16mm
\begin{center}
\mbox{\epsfig{file=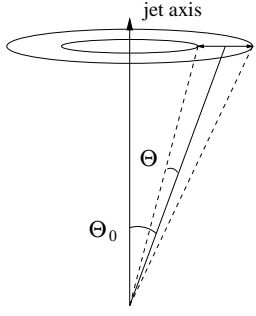,width=3.3cm}}
\end{center}
\vs-1cm
\caption{
A schematic representation of the interval in the polar angle around 
the jet axis. 
}
\label{pic}
\vs-4mm
\end{figure}

\medskip

1) In the fixed coupling regime, for  moderately small angular bins,
\beq
D_q=\gamma_0(Q)\frac{q+1}{q},
\label{an2}
\eeq
where $\gamma_0(Q)=\sqrt{2\,C_{\mathrm{A}}\as (Q)/\pi}$ is
the anomalous QCD dimension calculated at  $Q\simeq  E\Theta_0$,
$E=\sqrt{s}/2$, and 
gluon color factor $C_{\mathrm{A}}=N_{\mathrm{c}}=3$.

2) In the  running-coupling regime, for  small bins,
the  R\'{e}nyi dimensions  become  
a function of the size of the angular ring
($\as(Q)$ increases with decreasing  $\Theta$).
It is useful to introduce a new scaling variable \cite{brax}, 
\beq
z=\frac{\ln (\Theta_0/\Theta )}{\ln (E\Theta_0/\Lambda )},
\eeq
where the maximum possible region ($\Theta=\Theta_0$) corresponds to $z=0$.

Three approximations derived in DLLA  will be compared: 

a) According to \cite{drem}, the $D_q$ have the form 

\beq
D_q\simeq\gamma_0(Q)\frac{q+1}{q}\left(1 + \frac{q^2+1}{4q^2}z\right).
\label{an3}
\eeq

b) Another approximation has been suggested in \cite{brax}:

\beq
D_q\simeq2\,\gamma_0(Q)\frac{q+1}{q}\left(\frac{1-\sqrt{1-z}}{z}\right).
\label{an4}
\eeq

c) In \cite{ochs}, a result has 
been obtained for the cumulant moments converging to
factorial moments for high energies,

\beq
D_q\simeq2\,\gamma_0(Q)\frac{q-w(q,z)}{z(q-1)}, 
\eeq
\beq
w(q,z)=q\sqrt{1-z}\left(1-\frac{\ln(1-z)}{2q^2}\right).
\label{an44}
\eeq

In \cite{drem}, an  estimate  for  $D_q$  
has, furthermore, been obtained
in the Modified Leading Log Approximation (MLLA).
In this case,  (\ref{an3}) 
remains valid, but $\gamma_0(Q)$ is replaced by  an
effective $\gamma_0^{\mathrm{eff}}(Q)$ depending on $q$: 
\begin{eqnarray}
~\hskip-0.7cm & &\gamma_0^{\mathrm{eff}}(Q)=\gamma_0(Q) + 
\gamma_0^2(Q)\frac{b}{4C_{\mathrm{A}}} \ \cdot\hfill \nonumber \\
~\hskip-0.7cm & & \cdot\ \left[-B\frac{q-1}{2(q+1)} + \frac{q-1}{2(q+1)(q^2+1)} + 
\frac{1}{4}\right], \hfill
\label{an5}\\
~\hskip-0.7cm & & b=\frac{11C_{\mathrm{A}}}{3} - \frac{2n_{\mathrm{f}}}{3}, 
\qquad  B=\frac{1}{b}\left[\frac{11C_{\mathrm{A}}}{3} +
\frac{2n_f}{3C_{\mathrm{A}}^2}\right], \hfill \nonumber
\end{eqnarray}
where $n_{\mathrm{f}}$ is the number of flavors. 

For our comparison of data and 
theoretical predictions quoted above,
we will use $n_{\mathrm{f}}=3$ and $\Lambda=0.16\GeV$.

The low value of $n_{\mathrm{f}}$ is  chosen  
since even at high energies the production of higher flavors will rarely
happen in the jet and its 
evolution is still dominated by the light
flavors \cite{nf3}. (The contribution of heavy quarks
will be discussed  below.) 
The value of $\Lambda$ chosen is that found in tuning the JETSET 7.4 ME 
program \cite{sjo} 
on L3 data \cite{l3d} and in our most recent determination of
$\as (m_{\mathrm{Z}})$ \cite{alex}. 

For  the angle $\Theta_0$, we consider  
two possibilities: $\Theta_0=25^0 \quad  \mbox{and} \quad 35^0$.
The first value is chosen in order to compare our results  
with the DELPHI analysis \cite{del}, while the
larger value of $\Theta_0$ allows a larger range of $\Theta$ to be studied.

The effective coupling constant is evaluated at  
$Q\simeq E\Theta_0$. For $\Theta_0=25^0$, one obtains 
$\as(E\Theta_0)\simeq 0.144$ according  
to the first-order QCD expression for $\as (Q)$. 
This value leads to $\gamma_0(E\Theta_0)\simeq 0.525$.
For $\Theta_0=35^0$, $\as(E\Theta_0)\simeq 0.135$ and
$\gamma_0(E\Theta_0)\simeq 0.508$.

\subsection{The data}

The data as well as the selection and correction procedures are the
same as in Sect.~2.2, except that a more severe cut is applied on the track 
quality and $K^0_s$, $\Lambda^0$ decay products are excluded from the sample.
Furthermore, Bose-Einstein correlations and Dalitz decay are removed from 
the data using correction from MC.

The resolution of the L3 detector for 
a number of relevant variables 
was estimated in \cite{chekth}.
The resolution of polar angle defined
with respect to the thrust axis
is found to be approximately $0.01$  radians.
For higher-order NFMs,
the minimum angle  $\Theta$ is chosen according to 
the many-particle resolution studied in \cite{chekth}.

The error bars on the results include 
contributions from statistical errors on the raw quantities
and both statistical and systematic errors on the correction factors.

As a systematic error, we take half of the difference 
between the correction factors
determined using JETSET \cite{sjo} and those using HERWIG 5.9 \cite{marc}.
In addition, we checked how
a variation of the cuts affects  the NFMs. 
The influence  of such variations  on the observed signal
was  found to be negligible.

The sphericity axis is used to define the jet axis.

\subsection{Analysis}

\subsubsection{\mbox{Comparison~with~JETSET~on~ha-} dronic and partonic levels} 
\begin{figure}[tb]
\vs-15mm
\mbox{\epsfig{file=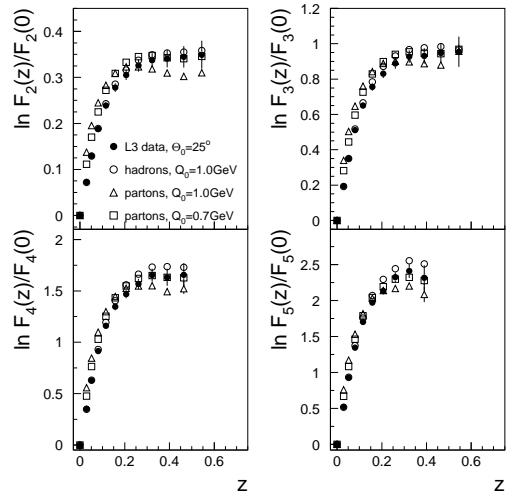,width=7cm}}
\vs-1cm
\caption{
The behavior of  $F_q(z)/F_q(0)$ ($q=2,\ldots, 5$) and
JETSET 7.4 PS predictions,  
on the partonic and hadronic levels.
Data are corrected for detector effects, initial-state photon radiation,
Bose-Einstein correlations and Dalitz decays.
}
\label{pic2}
\vs-5mm
\end{figure}

It has been suggested in \cite{brax}
to consider 
the ratio $F_q(z)/F_q(0)$ in order to reduce
hadronization effects on the actual behavior of the NFMs. 
In addition,
this   reduces  a theoretical ambiguity
for the evaluation  of   NFMs in  full phase space ($z=0$).
In terms of $F_q(z)/F_q(0)$, 
the power law (\ref{an1}) reads as 
\beq
\ln \frac{F_q(z)}{F_q(0)}=z(1-D_q)(q-1)\ln \frac{E\Theta_0}{\Lambda}.
\label{an6}
\eeq
The behavior of the $\ln (F_q(z)/F_q(0))$ as a function of $z$ is
shown in Fig.~\ref{pic2}. 

To increase statistics, we  evaluated
the NFMs in each  sphericity hemisphere  of an event 
and  averaged the results, thus  assuming
that the local fluctuations in each hemisphere  are independent. 

The open symbols show the predictions of 
the JETSET 7.4 PS  model for
hadronic (open circles) and 
partonic (open triangles) levels with
the default cut-off $Q_0=1.0\GeV$.
The Monte-Carlo  prediction for hadrons 
gives a reasonable description of the
fluctuations.

The partonic level of  JETSET is 
close to the data, both for  $Q_0=1.0\GeV$ and $Q_0=0.7\GeV$, though  
a noticeable  difference between 
the slopes for the data and the 
partonic level of  JETSET  is present.
The similarity of partonic and hadronic level Monte-Carlo
predictions is used as a measure of the degree of validity of LPHD.

The contributions of heavy flavors  (c and b quarks) were 
estimated by rejection  of these flavors on generator-level  
of JETSET. The  effect was found to be negligible.
 
\begin{figure}[tb]
\vs-1.5cm
\mbox{\epsfig{file=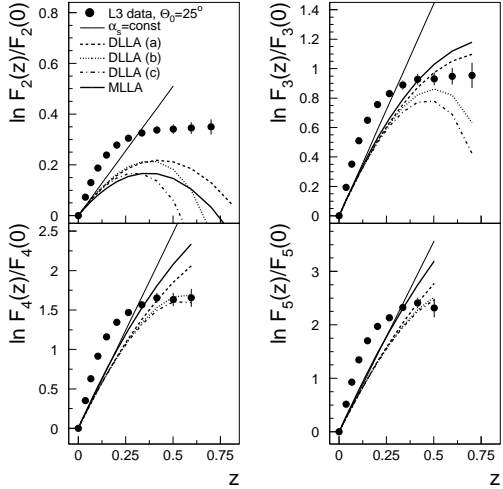, width=7cm}}
\vs-1cm
\caption{
The analytical QCD predictions for $\Lambda=0.16\>\mathrm{GeV}$:
$\as=\mathrm{const}$  (4);
DLLA (a) (6); DLLA (b) (7);  DLLA (c) (8);
MLLA (10).
}
\label{pic3}
\vs-5mm
\end{figure}

\subsubsection{Comparison of the analytical predictions}

The comparison of the analytical QCD predictions 
(\ref{an2})-(\ref{an5})
to the corrected data is shown in Fig.~\ref{pic3}  
for $\Lambda=0.16\GeV$ and  $\Theta_0=25^0$. 
For second order, the running-$\as$ predictions lead to the saturation
effects observed 
in the data, but significantly underestimate 
the observed signal.
Predictions for the higher moments are too low for low values of $z$, but tend
to overestimate the data at larger $z$.
The fixed coupling regime (thin solid lines) approximates 
the running coupling 
regime for small $z$, but does not exhibit 
the saturation effect seen in the data.
The DLLA approximation (\ref{an3}) and (\ref{an4}) differ 
significantly at large $z$.
The predictions for cumulants (8) 
show a stronger saturation effect. 
The MLLA predictions  do not  
differ significantly 
from the DLLA  result (\ref{an3}).

At $\Theta_0=35^0$ (not shown)  both data
and predictions show a stronger increase than for  $\Theta_0=25^0$. 
This indicates that fluctuations  are  larger for phase-space regions 
containing a larger contribution from hard-gluon radiation. 

\begin{figure}[tb]
\vs-15mm
\mbox{\epsfig{file=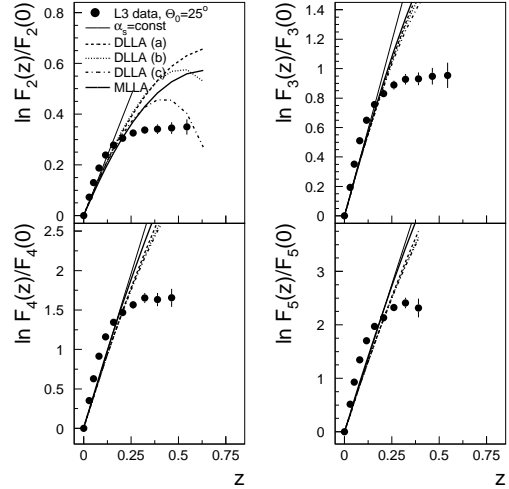, width=7cm}}
\vs-1cm
\caption{
Same as Fig.~10, but for $\Lambda=0.04\>\mathrm{GeV}$.
}
\label{pic5}
\vs-0.65cm
\end{figure}

A better agreement of  the QCD predictions  
with the data at low $z$-values
can be achieved by decreasing the value of $\Lambda$.
A similar observation has been made 
by  DELPHI \cite{del}.
As an example, Fig.~\ref{pic5} shows the case of $\Lambda=0.04\GeV$.
Such an effective value makes the coupling constant smaller and 
this  can expand  the range of reliability of the perturbative 
QCD calculations (for $\Lambda=0.04\GeV$, $\as(E\Theta_0)\simeq 0.112$,
$\gamma_0(E\Theta_0)\simeq 0.46$). 
However, this leads to a large disagreement 
between the QCD predictions 
and the data for $z>0.3$, where contributions from high-order
perturbative QCD and hadronization are expected to be stronger. 
We have varied $\Lambda$ in the range 
of $0.04-0.25\GeV$ and found that there is
no value of $\Lambda$ in this range which provides
agreement for all orders of NFMs.
Furthermore, increasing $n_{\mathrm{f}}$ 
to 4 and 5 reduces the second moment to
reproduce the data at large $z$, but 
high-order moments still continue to overshoot.

\subsection{Conclusions from local fluctuations}

The first-order predictions of the DLLA of  perturbative QCD are shown 
to be in disagreement with the local fluctuations as observed for 
hadronic $\Z$ decay. This conclusion is valid for standard  values 
of $\Lambda$ ($\Lambda=0.16\GeV$) as well as for small values 
($\Lambda =0.04\GeV$). In the latter case, a reasonable estimate for $z<0.3$ 
can be reached, consistent with the DELPHI conclusion \cite{del}.
However, our analysis shows that, in this case, the theoretical 
NFMs strongly overestimate the data for relatively large $z$ (small 
$\Theta$), where contribution from high-order perturbative QCD is  stronger.

The prediction of  MLLA shows very similar a result as DLLA. Note, however, 
that a full MLLA correction for the $z$-dependence of $D_q$ has not been 
obtained, so far.  The MLLA prediction  quoted above only modifies $\gamma_0$,
but the $z$-dependence of NFMs is parameterized by DLLA, which is only 
asymptotically correct. One large effect in the MLLA approximation is the 
angular recoil effect, which is important for small $z$. This effect can 
change the value of NFMs at $z=0$ and, hence, the absolute normalization 
of the NFMs \cite{prav}. Note that a recent study \cite{cons} of 
energy-conservation in triple-parton vertices shows that the 
energy-conservation constraint is indeed sizeable and leads to a stronger 
saturation effect. 

Another likely reason for their disagreement with the experimental 
data is the asymptotic character of the QCD predictions, corresponding to 
an infinite number of partons  in an event. Furthermore, the failure of the 
predictions can lie with the LPHD hypothesis, which is used to justify 
comparison of the predictions of perturbative QCD to hadronic data.

\section*{Acknowledgments}
This work is part of the research program of the ``Stichting voor
Fundamenteel Onderzoek der Materie (FOM)'', which
is financially supported by the ``Nederlandse Organisatie voor
Wetenschappelijk Onderzoek (NWO)''.
We acknowledge the effort of all engineers and
technicians who have participated
in the construction and
maintenance of the LEP machine and the L3
detector.

\raggedright

\end{document}